\documentclass[a4paper,10pt]{JHEP3}

\usepackage[latin1]{inputenc}
\usepackage[english]{babel}
\usepackage{textcomp}
\usepackage{amsfonts}
\usepackage{amsmath}
\usepackage{amssymb}
\usepackage{mathrsfs}
\usepackage[all]{xy}
\usepackage{cite} 
\usepackage{graphicx}
\usepackage{booktabs}
\usepackage{euscript}
\usepackage{rotating}
\usepackage{enumerate}
\usepackage{url}

\numberwithin{equation}{section}

\numberwithin{equation}{section}

\newtheorem{theorem}{Theorem}

\DeclareMathOperator{\diag}{diag}

\DeclareMathOperator{\Imm}{Im}

\DeclareMathOperator{\trace}{Tr}

\newcommand{\cc}{\mathbb{C}}

\newcommand{\rr}{\mathbb{R}}

\newcommand{\zz}{\mathbb{Z}}
\newcommand{\ZZ}{\mathbb{Z}}
\newcommand{\hh}{\mathbb{H}}
\newcommand{\HH}{\mathbb{H}}

\newcommand{\nn}{\mathbb{N}}
\newcommand{\FF}{{\bf F}}

\newcommand{\be}{\begin{equation}}
\newcommand{\ee}{\end{equation}}
\newcommand{\bes}{\begin{equation*}}
\newcommand{\ees}{\end{equation*}}
\newcommand{\beq}{\begin{eqnarray}}
\newcommand{\eeq}{\end{eqnarray}}
\newcommand{\beqs}{\begin{eqnarray*}}
\newcommand{\eeqs}{\end{eqnarray*}}
\newcommand{\bea}{\begin{align}}
\newcommand{\eea}{\end{align}} 
\newcommand{\beas}{\begin{align*}}
\newcommand{\eeas}{\end{align*}} 
\newcommand{\besp}{\begin{equation}\begin{split}}
\newcommand{\eesp}{\end{equation}\end{split}}
\newcommand{\besps}{\begin{equation*}\begin{split}}
\newcommand{\eesps}{\end{equation*}\end{split}}

\newcommand{\transp}[1]{{}^t\!#1}

\newcommand{\smaq}{\left[ \begin{smallmatrix}}
\newcommand{\smat}{\left( \begin{smallmatrix}}
\newcommand{\smcq}{\end{smallmatrix}\right]}
\newcommand{\smct}{\end{smallmatrix}\right)}
\newcommand{\smag}{\left \{ \begin{smallmatrix}}
\newcommand{\smcg}{\end{smallmatrix}\right \}}

\makeatletter
  \DeclareRobustCommand*\textsubscript[1]{%
    \@textsubscript{\selectfont#1}}
  \newcommand{\@textsubscript}[1]{%
    {\m@th\ensuremath{_{\mbox{\fontsize\sf@size\z@#1}}}}}
\makeatother

\date{\today}

\title{Classical theta constants vs. lattice theta series,\\
and super string partition functions}

\author{Francesco Dalla Piazza \\ Dipartimento di Fisica e Matematica, Universit\`a degli Studi dell'Insubria,
Via Valleggio 11, 22100 Como, Italy, and INFN, via Celoria 16, 20133, Milano, Italy. \\ E-mail address: f.dallapiazza@uninsubria.it}

\author{Davide Girola\\
Dipartimento di Fisica e Matematica, Universit\`a degli Studi dell'Insubria,
Via Valleggio 11, 22100 Como, Italy. \\
E-mail address: davide.girola@gmail.com}

\author{Sergio L. Cacciatori \\
\\ Dipartimento di Fisica e Matematica, Universit\`a degli Studi dell'Insubria,
Via Valleggio 11, 22100 Como, Italy, and INFN, via Celoria 16, 20133, Milano, Italy. \\ E-mail address: sergio.cacciatori@uninsubria.it}

\abstract{
Recently, various possible expressions for the vacuum-to-vacuum superstring amplitudes has been proposed at genus $g=3,4,5$.
To compare the different proposals, here we will present a careful analysis of the comparison between the two main
technical tools adopted to realize the proposals: the classical theta constants and
the lattice theta series. We compute the relevant Fourier coefficients in order to relate the two spaces. We will prove the
equivalence up to genus 4.
In genus five we will show that the solutions are equivalent modulo the Schottky form and coincide if we impose the vanishing of the cosmological constant.
}
\begin{document}


\section{Introduction}\label{ssmeas}
%
In the perturbative  approach, superstring theory can be formulated using the path integral formalism outlined by Polyakov. The
computation of amplitudes from first principles is an old problem in string theory and
finds its roots in the correct mathematical definition of the theory. In a series of papers D'Hoker and Phong have obtained the expression for the
genus two superstring vacuum-to-vacuum amplitude from direct path integral computation. This is a remarkable result. Moreover,
they have proved that the amplitude is slice independent, i.e. independent on the
parametrization of the even and odd moduli appearing in the path integral. Their solution satisfies the nonrenormalization
theorems, as expected in superstring theory. The amplitude is expressed in terms of modular forms of suitable weight defined on a genus two
Riemann surface. The measure $d\mu[\Delta]$ appearing in the integral of the amplitude splits in a holomorphic and an antiholomorphic
part. This is an essential feature to perform the GSO projection in order to eliminate the tachyon and make the theory stable. All the
computations are made explicitly for $g=2$, but the authors argued that the procedure should work at any genus $g$.

Following the conjecture (not yet proved) of D'Hoker and Phong one can assume that, as for $g=2$, the expression of the vacuum-to-vacuum
genus $g$ superstring amplitude has the general form:
\be \label{1punto1}
Z^g=\int_{{\mathcal{M}}_g} (\det \Imm \Omega)^{-5} \sum_{\Delta \Delta'} c_{\Delta \Delta'} d\mu [\Delta] (\Omega) \wedge \overline {d\mu [\Delta']
(\Omega)}
\ee
with
\begin{eqnarray*}
&& d\mu[\Delta] (\Omega) =d\mu_{BOS} (\Omega) \Xi_8 (\Delta),\\
&& \Delta=\left[ \begin{array}{c} a\\b \end{array} \right], \qquad\ a,b \in \mathbb{Z}_2^g,
\end{eqnarray*}
$d\mu_{BOS} (\Omega)$ being the well defined bosonic measure, $\Xi_8 [\Delta] (\Omega)$ are suitable equivariant modular forms,
$c_{\Delta\Delta'}$ are phases realizing the opportune $GSO$ projection, $\Omega$ is the period matrix of the genus $g$ Riemann surface and $M_g$
is its moduli space.
Note that the measure splits in holomorphic and antiholomorphic part $d\mu [\Delta] (\Omega)$ and $\overline {d\mu [\Delta'] (\Omega)}$ respectively.
As discussed by Morozov, there are two different approaches to deal with the superstring measure. The first attitude is to try to prove
the general position \eqref{1punto1} by starting from the Polyakov's measure for $NSR$ string for a fixed
characteristic $\Delta$ and integrating out the odd moduli. In this way one would directly obtain the explicit expressions
for $d\mu[\Delta]$. This requires to realize the holomorphic factorization. In this procedure there are several subtle
points, not yet solved, due to the dependence of the result on the choice of the parametrization of even and odd moduli.
The second viewpoint is to assume the validity of \eqref{1punto1}, and, from general considerations, to make a reasonable guess for the measure
$d\mu[\Delta]$ and use its proprieties to determine it explicitly. In \cite{CDG,CDG2} it was adopted this second approach,
and, by a slightly modification of the ans\"atze of D'Hoker and Phong \cite{DP1,DP2} for the properties of $d\mu$, an expression for the
superstring chiral measure was found for $g\leq 4$. In the same spirit, in \cite{Grr} it has been proposed a
candidate for the $g$ loop measure, but, due to the presence of square and higher roots, it may be not well defined for $g\geq 5$.
Indeed, Salvati Manni in \cite{SM} discusses the case $g=5$ and shows that all the functions
appearing in the expression of the (candidate) five loop measure are well defined, at least on the moduli space of curves. In \cite{D} another expression for the five loop superstring measure was proposed using the classical theta constants in which does not appear any root. In all these
constructions the forms $\Xi_8[\Delta]$ are built up by making use of the classical theta constants with characteristic.
Following this approach one can state a general guess for the supersymmetric invariant measure
at any genera. Indeed, we  require some constraints for the functions $\Xi_8 [\Delta^{(g)}](\Omega^{(g)})$, which consist in three items
(see, for details, \cite{CDG,CD2}):
\begin{enumerate}[i.]
\item
The functions $\Xi_8^{(g)}[\Delta^{(g)}](\Omega^{(g)})$ are holomorphic on the Siegel upper halfplane\footnote{In the following we will not  always write the dependence of $\Xi_8^{(g)}$ on $\Omega^{(g)}$.} $\HH_g$ (regularity constraint).
\item
Under the action of the symplectic group $Sp(2g,\ZZ)$ on $\HH_g$, they should transform as follows (transformation constraint):
\begin{equation}\label{trasxi}
\Xi_8^{(g)}[M\cdot\Delta^{(g)}](M\cdot\Omega)\,=\, \det(C\Omega+D)^8\Xi_8^{(g)}[\Delta^{(g)}](\Omega),
\end{equation}
for all $M\in Sp(2g,\ZZ)$. Here, the affine action of $M$ on the characteristic $\Delta^{(g)}$ is given by
\be \label{actonchar}
\begin{pmatrix}A&B\\C&D\end{pmatrix}\cdot [{}^a_b]\,:=\,
[{}^c_d],\qquad
\left(\begin{array}{c}\transp{c}\\ \transp{d}\end{array}\right)\,=\,
\left(\begin{array}{cc} D&-C\\-B&A\end{array}\right)
\left(\begin{array}{c}\transp{a}\\ \transp{b}\end{array}\right)
\,+\,
\left(\begin{array}{c}(C\transp{D})_0\\(A\transp{B})_0\end{array}\right)
\quad \mbox{mod}\;2
\ee
where $N_0=(N_{11},\ldots,N_{gg})$ is the diagonal of the matrix $N$.
\item
The restriction of these functions to ``reducible" period matrices is a product of the corresponding
functions in lower genus (factorization constraint). More precisely, let
$$
D_{k,g-k}\,:=\,\left\{\Omega_{k,g-k}\,:=\,
\begin{pmatrix}\Omega_k&0\\0&\Omega_{g-k}\end{pmatrix}\,\in\HH_g\,:\,
\Omega_k\in \HH_k,\;\Omega_{g-k}\in\HH_{g-k}\,\right\}\;\cong\;
\HH_k\times\HH_{g-k}.
$$
Then we require that for all $k$, $0<k<g$,
$$
\Xi_8^{(g)}[{}^{a_1\ldots a_k\,a_{k+1}\ldots a_g}_{b_1\ldots b_k\,b_{k+1}\ldots b_g}](\Omega_{k,g-k})\,=\,
\Xi_8^{(k)}[{}^{a_1\ldots a_k}_{b_1\ldots b_k}](\Omega_k)
\Xi_8^{(g-k)}[{}^{a_{k+1}\ldots a_g}_{b_{k+1}\ldots b_g}](\Omega_{g-k})
$$
for all even characteristics
$\Delta^{(g)}=[{}^{a_1\ldots a_g}_{b_1\ldots b_g}]$ and all
$\Omega_{k,g-k}\in D_{k,g-k}$.
\end{enumerate}
One would hope that these constraints characterize uniquely the
measure. Indeed, in \cite{CDG,CDG2,OPSY} the uniqueness of the form $\Xi_8^{(g)}[\Delta^{(g)}]$ has been proved for $g\leq4$.

In \cite{OPSY} another candidate for the genus five superstring measure has been
proposed. The authors have made use of the notion of lattice theta series, see Section \ref{lts}. The forms $\Xi_8[\Delta]$ defined there
to build up the measure, just like the ones obtained using the classical theta constants, satisfies all the constraints.
The same formalism has been used to obtain the expressions of the measures for $g\leq 4$. It is not
clear if the two constructions are equivalent and lead to the same forms $\Xi_8[\Delta]$, thereby to the same measure
$d\mu[\Delta]$. Obviously, this is the case for $g\leq 4$, as a consequence of the uniqueness theorems in lower genus. The goal of
the present paper is to show that also in genus five the two constructions are equivalent, and the forms obtained are equal on the whole Siegel
half upper plane $\hh_5$, provided we add to the three constraints the supplementary request of vanishing cosmological constant. Otherwise they could
differ for a multiple of the Schottky form $J^{(5)}$ (that vanishes on the locus of trigonal curves, cf. \cite{GS}).
Actually, adding a scalar multiple of $J^{(5)}$ to a form satisfying the three constraints one obtains a function again satisfying the same constraints: the Schottky $J^{(5)}$ is a modular form of weight eight and the restriction to $\hh_1\times \hh_4$ is proportional to $F_{16}^{(1)}$ times $J^{(4)}$ and this product vanishes on the Jacobi locus. This is a remarkable fact because, differently from the genus four case, the zero locus $J^{(5)}$ is not the whole Jacobi locus, but the space of trigonal curves. Thus, the three constraints do not characterize uniquely the superstring measure, see \cite{GS,DbMS,MV}. This freedom can be fixed requiring the vanishing of the cosmological constant. Nevertheless, this should be a prediction of the theory and it should not be imposed by hand.
This is a remarkable result both for the viewpoint of physics and of mathematics. Indeed, this shows that there are an infinity of
different forms satisfying the three constraints on $\hh_5$, actually on $J_5$. Thus, the constraints, without the additional request on the cosmological constant, do not suffice to characterize the measure uniquely in any
genus. Furthermore, a deeper question arises about the conjecture by D'Hoker and Phong on the general expression \eqref{1punto1} for the
superstring chiral measure and about the procedure leading to it. These issues are at the basis of the mathematical correct
formulation of the string theory in the perturbative approach. To solve these problems some more insight in the physics leading to the
(conjectured) ansatz \eqref{1punto1} is necessary.

Mathematically, to prove the equivalence of the forms $\Xi_8^{(5)}$, one has to show that the space spanned by the lattice theta series and the one spanned by the eight functions defined in
Section \ref{rtc} (that are a basis for $M_8^\theta(\Gamma_5(2))$, cf. \cite{D}) are the same space of dimension eight.
This is the content of the following theorem:
\begin{theorem} \label{teo1}
The spaces $M_8^\theta(\Gamma_5(2))^{O^+}$ and $M_8^{\Theta_S}(\Gamma_5(2))$ coincide.
\end{theorem}
Here $M_8^\theta(\Gamma_5(2))^{O^+}$ is the space of genus five modular forms of weight
eight with respect to the group $\Gamma_5(2)$ that are $O^+$-invariant polynomials in the classical theta constants,
$M_8^{\Theta_S}(\Gamma_5(2))$ is the space of modular forms of weight eight spanned by the lattice theta series
($[\Gamma_5^{\Theta_S}(1,2),8]$ in the notation of \cite{OPSY}), and $M_8(\Gamma_5(2))^{O^+}$
is the space of genus five modular forms of weight eight with respect to the group $\Gamma_5(2)$, which are $O^+$-invariant,
cf. \cite{CD2, DG,D,OPSY,MV} for details.
The theorem follows from a result of Salvati-Manni\footnote{We are grateful to Riccardo Salvati-Manni who has put his papers to our attention, and explained his main theorem to us.} \cite{SM2,SM3,SM4} in which it was proved that the space generated by the lattice theta series contains the subspace  generated by classical theta constants that are $\Gamma_g$  invariant whenever 4 divides the weight (see also \cite{freitag}, theorem VI.1.5). The result applies also for the $\Gamma_g(1,2)$ case and, as a consequence, one has:
\be
M^{\theta}_{4k}(\Gamma_g(2))^{O^+}\subset   M^{\Theta_S}(\Gamma_g(2)),
\ee
for integer $k$.
In genus five the dimensions of both spaces is eight, see \cite{OPSY} for the $M^{\Theta_S}(\Gamma_g(2))$ case, and \cite{D} for the $M^{\theta}_{8}(\Gamma_g(2))^{O^+}$ one where also a basis for this space has been constructed. 
Thus, the theorem follows from the equality of the dimensions of the spaces.

In this paper we exhibit a complete map between the two spaces obtaining all the linear relations between the lattice theta series and the basis functions of the space  $M^{\theta}_{8}(\Gamma_5(2))^{O^+}$ defined in Section \ref{rtc} by means of the classical theta constants.
To obtain the map we compute certain Fourier coefficients of the functions appearing in the definition of the superstring
measure. Since the spaces $M_8^\theta(\Gamma_5(2))^{O^+}$ and
$M_8^{\Theta_S}(\Gamma_5(2))$ have dimension eight (see \cite{D,OPSY}) we need at least eight suitable
Fourier coefficients to get linear isomorphisms between these spaces and two copies of $\cc^8$.
In particular, being the two spaces the same, there must be linear relations among the
Fourier coefficients of the elements of the two bases, which obviously extend to the complete series.
In Section \ref{analproof} we also give an analytic proof of the
equivalence between the functions $\Xi_8[\Delta]$ constructed employing the three constraints and the supplementary request of the vanishing of the cosmological constant. In addition, the Fourier coefficients method will permit to obtain,
for $g\leq 4$, the complete set of linear relations between the
lattice theta series and the basis functions of $M_8^\theta(\Gamma_g(2))^{O^+}$. We will also check the well known linear relations
among the lattice theta series themselves \cite{DbMS,OPSY}.

\section{Lattice theta series} \label{lts}
\subsection{Lattices and theta series}
In this section we review the notion of lattices, quadratic forms associated with them and lattice theta series,
see \cite{AZ,CS} for details.
An $n$ dimensional lattice in $\rr^n$ has the form $\Lambda=\{\sum_{i=1}^n a_iv_i\; {\rm s.t.}\; a_i\in\zz\}$, where $v_i$ are
the elements of a basis of $\rr^n$ and are called basis for the lattice. A
fundamental region is a building block which when repeated many times
fills the whole space with just one lattice point in each copy.
Different basis vector could define the same lattice,
but the volume of the fundamental region is uniquely determined by
$\Lambda$. The square of this volume is called the determinant or
discriminant of the lattice.
The matrix
\be
M=
\begin{pmatrix}
v_{11} & \cdots & v_{1m} \\
\vdots & & \vdots \\
v_{n1} & \cdots & v_{nm}
\end{pmatrix},
\ee
where $v_i=(v_{i1},\cdots,v_{im})$ are the basis vectors is called
generator matrix for the lattice. The matrix $A=M\transp{M}$
is called Gram matrix and the entry $(i,j)$ of $A$ is the inner product
$v_i\cdot v_j$. The determinant of $\Lambda$ is the determinant of
$A$.
A generic vector $x=(x_1,\cdots,x_n)$ of the lattice can
be written as $x=\zeta M=\zeta_1v_1+\cdots + \zeta_nv_n$, where $\zeta=(\zeta_1,\cdots,\zeta_n)$ is an
arbitrary vector with integer components. Its norm is
$N(x) =x\cdot x=\zeta A \transp{\zeta}$. This is a quadratic form associated with the lattice in the integer variables $\zeta_1,\cdots,\zeta_n$.
Any $n$-dimensional lattice $\Lambda$ has a dual lattice, $\Lambda^*$,
given by:
\be
\Lambda^*=\{x\in\rr^n\,\,s.t.:\,\,x\cdot u\in\zz\,\,\mbox{for all}\,\,
u\in\Lambda\}.
\ee

If a lattice can be obtained from another one by a rotation, reflection
and change of scale we say that the two lattices are equivalent (or similar). Two
generators matrices define equivalent lattices if and only if they are
related by $M'=c\,UMB$, where $c$ is a non zero constant, $U$ is a
matrix with integer entries and determinant $±1$, and $B$ is a real
orthogonal matrix. Then, the corresponding Gram matrices are related by
$A'=c^2UA\transp{U}$. If $c=1$ the two lattices are congruent and if
also $\det U=1$ they are directly congruent.
Quadratic forms corresponding to congruent lattices are called
integrally equivalent, so there is a one to one correspondence between
congruence classes of lattice and integral equivalence classes of
quadratic forms. If $\Lambda$ is a lattice in $n$-dimensional space
that is spanned by $n$ independent vectors (i.e. a full rank lattice),
then $M$ has rank $n$, $A$ is a positive definite matrix, and the
associated quadratic form is called a positive definite form.
A lattice or a quadratic form is called integral if the inner product
of any two lattice vectors is an integer or, equivalently, if the Gram
matrix $A$ has integer entries. One can prove that a lattice is
integral if and only if $\Lambda\subseteq \Lambda^*$. An integral
lattice with $\det\Lambda=1$, or equivalently with
$\Lambda=\Lambda^*$ is called unimodular or self-dual. If $\Lambda$ is
integral then the inner product $x\cdot x$ is necessarily an integer
for all points $x$ of the lattice. If $x\cdot x$ is an even
integer for all $x\in\Lambda$ then the lattice is called even,
otherwise odd. Even unimodular lattices exist if and only if the
dimension is a multiple of 8, while odd unimodular lattices exist in
all dimensions.

For a lattice $\Lambda$ let $N_m$ be the number of vectors
$x\in\Lambda$ of norm $m=x\cdot x$. Thus, $N_m$ is also the number of
integral vectors $\zeta$ that are solutions of the Diophantine equation
\be
\zeta A \transp{\zeta} =m
\ee
or, in other words, the number of times that the quadratic form associated with $\Lambda$ represents the number $m$.
The (genus one) theta series of a lattice $\Lambda$ is a holomorphic
function on the Siegel upper half space $\hh_1$, defined by
\be
\Theta_\Lambda(\tau)=\sum_{x\in\Lambda}q^{x\cdot
  x}=\sum_{m=0}^\infty N_mq^m,
\ee
where $q=e^{\pi i \tau}$ and $\tau\in\hh_1$.
For example, the theta series associated to the lattice $\zz$ is the
classical Jacobi theta constant $\Theta_{\zz}(\tau)=\sum_{m=-\infty}^{\infty}q^{m^2}=1+2q+2q^4+2q^9+\cdots
\equiv\theta[{}_0^0](\tau)$, see Section \ref{rtc}.
This definition generalizes to theta series of arbitrary genus $g$. In
this case the vector $\zeta$ becomes a $g\times n$ matrix $\underline{\zeta}$ with integer
entries. In addition, one also introduces a $g\times n$ array $\underline{x}$
whose rows are the vectors of the lattice $\Lambda$. It can be written as  $\underline{x}=\underline{\zeta}M$.
Let $N_{\underline{m}}\in\zz$ be the number of integral matrix solutions of the Diophantine system
\be
\underline{\zeta} A \transp{\underline{\zeta}} =\underline{m},
\ee
where $\underline{m}$ is a $g\times g$ symmetric matrix whit integer entries. The
component $(i,j)$ of $\underline{m}$ represents the scalar product between the vectors
$x_i\in\Lambda$ and $x_j\in\Lambda$ of  $\underline{x}$. Thus,
$N_{\underline{m}}$ is also the number of the sets $\underline{x}$ of
$g$-vectors such that $x_i\cdot x_j=m_{ij}$.
In the same spirit of the genus one case, the genus $g$ theta series
associated to a lattice $\Lambda$ is a holomorphic function on the Siegel upper half space $\hh_g$, defined by
\be \label{thserg}
\Theta_\Lambda^{(g)}(\tau)=\sum_{x\in\Lambda^{(g)}}e^{\pi i\trace(x\cdot
  x)}=\sum_{\underline{\zeta}\in\zz^{g,n}}e^{\pi i\trace(\underline{\zeta} A \transp{\underline{\zeta}}\tau)}
  =\sum_{\underline{m}} N_{\underline{m}}\prod_{i\leq j}e^{\pi i m_{ij}\tau_{ij}},
\ee
and $\tau\in\hh_g$. Lattice theta series corresponding to a self-dual $n$-dimensional
lattice, with $n$ divisible by 8, is a modular form of weight $\frac n2$ with respect to the group $\Gamma_g(1,2)$ if the lattice is odd
and with respect to $\Gamma_g$ if the lattice is even\footnote{We recall the definitions:
\begin{align*}
& \Gamma_g:={\rm Sp}(2g,\mathbb{Z}), \qquad \Gamma_g(2):=\{ M\in \Gamma_g \mid M=\mathbb{I}\ {\rm mod} 2 \} \\
& \Gamma_g (1,2) :=\{ M=\left(^A_C{}^B_D \right) \in \Gamma_g (2) \mid \ A \ \transp B ={\rm diag} C\ \transp D =0 \ {\rm mod} 2 \}.
\end{align*}
}. Thus, lattice theta series associated to 16-dimensional self-dual lattices are
modular forms of weight 8. There are eight 16-dimensional self-dual
lattice \cite{CS}, two even and six odd, and they can be obtained from the root lattice of
some Lie algebra. See also \cite{DbMS, OPSY,MV}.
In what follows we will use a nice property of lattice theta series
when restricted to block diagonal period matrices: indeed, they factorize in a very simple way when $\tau\in\hh_k\times\hh_{g-k}$:
\be \label{factts}
\Theta_\Lambda^{(g)}
\begin{pmatrix}
\tau_k & 0 \\
0 & \tau_{g-k}
\end{pmatrix}
=\Theta_\Lambda^{(k)}(\tau_k) \Theta_\Lambda^{(g-k)}(\tau_{g-k}).
\ee

\subsection{Fourier coefficients of lattice theta series} \label{flts}
In order to express the relations between lattice theta series and the
classical theta constants, we first expand in Fourier series the
lattice theta constants. We just need the coefficient
$N_{\underline{m}}$ of the series \eqref{thserg} for some integer
matrix $\underline{m}$. It is known (cf. \cite{OPSY}) that in genus
five the eight theta series are all independent, whereas for lower
genus there are linear relations among them. Thus, we have to choose
at least eight $\underline{m}$ in such a way that the matrix of the
Fourier coefficients $N_{\underline{m}}$ of the eight theta series has rank 8.
In Table \ref{tab:fouriercoef} are shown the Fourier coefficients for
the eight theta series up to $g=5$. We computed the coefficients for
the matrices:
\begin{align*}
m_1&=\smat
1 & 0 & 0 & 0 & 0 \\
0 & 0 & 0 & 0 & 0 \\
0 & 0 & 0 & 0 & 0 \\
0 & 0 & 0 & 0 & 0 \\
0 & 0 & 0 & 0 & 0 \\
\smct
&m_2&=\smat
2 & 0 & 0 & 0 & 0 \\
0 & 0 & 0 & 0 & 0 \\
0 & 0 & 0 & 0 & 0 \\
0 & 0 & 0 & 0 & 0 \\
0 & 0 & 0 & 0 & 0 \\
\smct
&m_3&=\smat
3 & 0 & 0 & 0 & 0 \\
0 & 0 & 0 & 0 & 0 \\
0 & 0 & 0 & 0 & 0 \\
0 & 0 & 0 & 0 & 0 \\
0 & 0 & 0 & 0 & 0 \\
\smct
&m_4&=\smat
1 & 0 & 0 & 0 & 0 \\
0 & 1 & 0 & 0 & 0 \\
0 & 0 & 0 & 0 & 0 \\
0 & 0 & 0 & 0 & 0 \\
0 & 0 & 0 & 0 & 0 \\
\smct
&m_5&=\smat
2 & 0 & 0 & 0 & 0 \\
0 & 2 & 0 & 0 & 0 \\
0 & 0 & 0 & 0 & 0 \\
0 & 0 & 0 & 0 & 0 \\
0 & 0 & 0 & 0 & 0 \\
\smct \\
m_6&=\smat
1 & 0 & 0 & 0 & 0 \\
0 & 1 & 0 & 0 & 0 \\
0 & 0 & 1 & 0 & 0 \\
0 & 0 & 0 & 0 & 0 \\
0 & 0 & 0 & 0 & 0 \\
\smct
&m_7&=\smat
2 & 0 & 0 & 0 & 0 \\
0 & 2 & 0 & 0 & 0 \\
0 & 0 & 2 & 0 & 0 \\
0 & 0 & 0 & 0 & 0 \\
0 & 0 & 0 & 0 & 0 \\
\smct
&m_8&=\smat
1 & 0 & 0 & 0 & 0 \\
0 & 1 & 0 & 0 & 0 \\
0 & 0 & 1 & 0 & 0 \\
0 & 0 & 0 & 1 & 0 \\
0 & 0 & 0 & 0 & 0 \\
\smct
&m_9&=\smat
2 & 0 & 0 & 0 & 0 \\
0 & 2 & 0 & 0 & 0 \\
0 & 0 & 2 & 0 & 0 \\
0 & 0 & 0 & 2 & 0 \\
0 & 0 & 0 & 0 & 0 \\
\smct
&m_{10}&=\smat
1 & 0 & 0 & 0 & 0 \\
0 & 1 & 0 & 0 & 0 \\
0 & 0 & 1 & 0 & 0 \\
0 & 0 & 0 & 1 & 0 \\
0 & 0 & 0 & 0 & 1 \\
\smct.
\end{align*}
Appealing to the geometric interpretation for the matrices
$\underline{m}_{k}$, with $k=1,\cdots, 10$,
for each of the eight 16-dimensional self-dual lattices, we
are looking for the number of integer solutions of the Diophantine
equation $\underline{\zeta}A_\Lambda\transp{\underline{\zeta}}=\underline{m}_k$. In
other terms, we are counting the number of sets $\underline{x}$ of five vectors in the
lattice $\Lambda$ such that the vector $x_i$ has norm $(m_k)_{ii}$ and the
inner product with the vector $x_j$ is $x_i\cdot x_j=(m_k)_{ij}$.
It is clear that the Fourier coefficients corresponding, for example,
to the matrix $m_4$ can be interpreted as the Fourier coefficients
for the genus two theta series in which the two orthogonal vectors
$x_1$ and $x_2$ have both norm 1, but also as the coefficients
of the theta series of genus $g>2$ in which the vectors
$x_i$ with $i>2$ have null norm.
It is not hard to perform this computation using a software like
Magma, although the computation of the coefficients corresponding to
the matrix $\diag(2,2,2,2,0)$ may take some hours.

The notation of the Table \ref{tab:fouriercoef} is the same as in
\cite{OPSY}. The rows contain the Fourier coefficients of the
theta series corresponding to the eight lattices $(D_8\oplus D_8)^+$,
$\zz\oplus A_{15}^+$, $\zz^2\oplus (E_7 \oplus
  E_7)^+$, $\zz^4\oplus D_{12}^+$, $\zz^8\oplus E_8
$, $\zz^{16}$, $E_8\oplus E_8$, and $D_{16}^+$, where the last two
lattices are the even ones. At the top of the columns we just
indicated the diagonal elements of the matrices $\underline{m}_k$, the other
elements being zero.
As anticipated, the rank of the full matrix of the coefficient is
eight, thus no linear relations between genus five theta series
exist. However, considering the same matrix for genus less than
five one can obtain the relations between theta series, as we will
show in the following, for every $g\leq 4$. In the top of the Table we write in
bold the matrices strictly necessary for the computation, whereas some other columns are added as a check.
The same convention will be used throughout in the paper.

\begin{table}[hbtp]
\begin{center}
\resizebox*{1\textwidth}{!}{
\begin{tabular}{lcccccccccc}
\toprule
&$ {\bf (1,0,0,0,0)} $&${\bf (2,0,0,0,0)} $&$ {\bf (3,0,0,0,0)} $&$
{\bf (1,1,0,0,0)} $ &$ (2,2,0,0,0) $ &$ {\bf (1,1,1,0,0)} $ &$ (2,2,2,0,0) $ &$ {\bf (1,1,1,1,0)} $&$ {\bf (2,2,2,2,0)} $&$ {\bf (1,1,1,1,1)}$ \\
\midrule
$\Theta_{(D_8\oplus D_8)^+}$   & 0    &  224  & 4096   & 0     & 38976
& 0     & 5069568 &  0        & 475270656 & 0 \\
$\Theta_{\zz\oplus A_{15}^+}$  &$ 2  $&$ 240 $&$ 4120 $&$ 0 $  & 43680
&$ 0 $  & 5765760 &  $ 0$     & 518918400 & 0\\
$\Theta_{\zz^2\oplus (E_7 \oplus E_7)^+}$ &$ 4  $&$ 256 $&$ 4144 $&$ 8
$  & 48896 &$ 0 $  & 6676992 & $ 0$      & 644668416 & 0\\
$\Theta_{\zz^4\oplus D_{12}^+}$ &$ 8  $&$ 288 $&$ 4192 $&$ 48 $ &
60864 &  192  & 9181440 &$ 384$     & 964200960 & 0\\
$\Theta_{\zz^8\oplus E_8} $  &$ 16 $&$ 352 $&$ 4288 $&$ 224 $ & 90944
& 2688  & 17176320 & $ 26880 $ & 2316142080 & 215040\\
$\Theta_{\zz^{16}}$   &$ 32 $&$ 480 $&$ 4480 $&$ 960 $& 175680 & 26880
& 47174400 & $ 698880 $& 8858304000 & 16773120\\
$\Theta_{E_8\oplus E_8}$   &$ 0  $&$ 480 $&$ 0 $   &$ 0 $  & 175680 &
$ 0 $  & 47174400 &  $ 0$     & 9064742400 & 0\\
$\Theta_{D_{16}^+}$  &$ 0  $&$ 480 $&$ 0 $   &$ 0 $  & 175680 & $ 0 $
& 47174400 & $ 0$      & 8858304000 & 0\\
\bottomrule
\end{tabular}
}
\caption{Fourier coefficients for the lattice theta series.}
\label{tab:fouriercoef}
\end{center}
\end{table}

\section{Riemann theta constants and the forms $\Xi_8^{(g)}$} \label{rtc} 
The form $\Xi_8^{(g)}[0^{(g)}]$, appearing in the expression for the superstring chiral measure, belongs to $M_8(\Gamma_g(2))^{O^+}$,
the space of modular forms of weight eight with respect to the group $\Gamma_g(2)$, and invariant under
the action of $O^+:=\Gamma_g(1,2)/\Gamma_g(2)$, see \cite{DG,D} for
details. In \cite{D} a basis for these spaces has been found for $g\leq 5$ and a suitable linear combination among these basis vectors has been
obtained by imposing the constraints of Section \ref{ssmeas}.

Let us now discuss the theta constants with characteristics, which are a powerful tool for constructing
modular forms on $\Gamma_g(2)$. An even characteristic is a $2\times
g$ matrix $\Delta=\left[{}^a_b\right]$, with $a,\,b\in\{0,1\}$  and
$\sum a_ib_i\equiv 0$ mod $2$. Let $\tau\in\hh_g$, the Siegel upper half space,
then we define the theta constants with characteristic:
\be
\theta[{}^a_b](\tau)\,:=\,\sum_{\transp m\in\zz^g} \,e^{\pi i((m+a/2)\tau\transp{(m+a/2)}+(m+a/2)\transp{b})},
\ee
where $m$ is a row vector. Thus, theta constants are holomorphic functions on $\hh_g$. One can build modular forms of weight eight as suitable polynomials of degree sixteen in the theta constants, see \cite{DG,D,CD2} for details.
Defining the $g\times g$ symmetric matrix $M$ with entries
$M_{ii}=m_i^2+a_im_i+\frac{a_i^2}{4}$, $i=1,\cdots,g$ and
$M_{ij}=m_i m_j+\frac{a_j}{2}m_i+\frac{a_i}{2}{m_j}+\frac{a_ia_j}{4}$,
$1\leq i<j\leq g$, the definition of theta constant can be rewritten as
\begin{align}\label{sum1}
\theta[{}^a_b](\tau)\,:&=\,\sum_{m\in\zz^g} \,
(-)^{\frac{a_1b_1}{2}+\cdots +\frac{a_gb_g}{2}}(-)^{b_1m_1+\cdots +
  b_gm_g}e^{\pi i\trace(M\tau)} \cr
&=(-)^{\frac{a_1b_1}{2}+\cdots +\frac{a_gb_g}{2}}\,\sum_{m\in\zz^g} \,
(-)^{b_1m_1+\cdots +
  b_gm_g}\prod_{i\leq j}e^{\pi i (2-\delta_{ij})M_{ij}\tau_{ij}} \cr
&=\,\sum_{A\in\frac{\zz^{g,g}}{4},\transp{A}=A}N_{A}\prod_{i\leq j}e^{\pi i A_{ij}\tau_{ij}},
\end{align}
where $A$ is a symmetric $g\times g$ matrix with entries in $\frac 14\zz$ and $N_A$ is an integer coefficient. In particular $N_A$ is the
number of times\footnote{Counted with signs given by the factor multiplying the product of exponentials.} that the particular matrix $A$
appears in the sum \eqref{sum1}. Note that the factor $(-)^{\frac{a_1b_1}{2}+\cdots +\frac{a_gb_g}{2}}$ is a global sign
depending only on the characteristic $\Delta$ and the coefficient $(-)^{b_1m_1+\cdots+ b_gm_g}$ is a sign depending on the second row
of the theta characteristic and on the matrix $M$.

In \cite{D} there have been computed the dimensions of the spaces of $O^+$-invariants for $g\leq 5$. It turned out
that these dimensions are $3$, $4$, $5$, $7$ and $8$ for $g=1$, $2$,
$3$, $4$ and $5$ respectively. Moreover, a basis has been provided for each of these genera, by means of the classical Riemann theta
constants. For each genus one defines:
\begin{align*}
F_1^{(g)} & :=\,\theta[0^{(g)}]^{16}, & F_8^{(g)} & :=\,(\sum_{\Delta^{(g)}}\theta[\Delta^{(g)}]^8)^2, \\
F_2^{(g)} & :=\,\theta[0^{(g)}]^4\sum_{\Delta^{(g)}} \,\theta[\Delta^{(g)}]^{12}, &F_{88}^{(g)} &
:=\,\sum_{(\Delta_i^{(g)},\Delta_j^{(g)})_o}\theta[\Delta_i^{(g)}]^8\theta[\Delta_j^{(g)}]^8,\\
F_3^{(g)} & :=\,\theta[0^{(g)}]^8\sum_{\Delta^{(g)}}\theta[\Delta^{(g)}]^8, &F_{16}^{(g)} & :=\, \sum_{\Delta^{(g)}}\theta[\Delta^{(g)}]^{16},
\end{align*}
where $(\Delta_i^{(g)},\Delta_j^{(g)})_o$ stands for the set of all pairs of distinct even characteristics whose sum
is odd.
Behind these, in \cite{D} are defined the forms $G_3^{(g)}[0^{(g)}]$ for $g=4,5$ and $G_4^{(g)}[0^{(g)}]$ for $g=5$. However,
we note that $G_3^{(g)}[0^{(g)}]$
($G_4^{(g)}[0^{(g)}]$) could be defined for every genus $g\geq 3$ ($g\geq 4$) considering three (four) dimensional isotropic subspace of
$\FF_2^{2g}$, where $\FF_2$ is the field of two elements. See \cite{CDG, CDG2, D, Grr} for more definitions and details.
Consider the form $J^{(g)}:=\,
2^g\sum_{\Delta^{(g)}}\theta[\Delta^{(g)}]^{16}-(\sum_{\Delta^{(g)}}\theta[\Delta^{(g)}]^8)^2=2^gF_{16}^{(g)}-F_8^{(g)}$. It
vanishes identically in genus $g\leq 2$, for $g=3$ vanishes on the
whole $\hh_3$, for $g=4$ on the Jacobi locus and for $g=5$ on the locus of trigonal curves.
Clearly all these functions are not linear independent for $g<5$, thus
for each genus we extract a basis as reported in Table \ref{tab:Obasis}, where the symbol $\surd$ means that the same function
as in lower genus has been taken as element of the basis (with obvious modifications).
\begin{table}[!h]
\begin{center}
\begin{tabular}{cccccc}
\toprule
Basis/$g$ &
1 & 2 & 3 & 4 & 5 \\
\midrule
$F_1$ & $\theta[0]^{16}$ & $\surd$ & $\surd$ & $\surd$ & $\surd$ \\
$F_2$ & $\theta[0]^4\sum_{\Delta} \,\theta[\Delta]^{12}$ & $\surd$ & $\surd$ & $\surd$ & $\surd$ \\
$F_{16}$ & $\sum_{\Delta}\theta[\Delta]^{16}$  & $\surd$ & $\surd$  & $\surd$  & $\surd$  \\
$F_3$ && $\theta[0]^8\sum_{\Delta}\theta[\Delta]^8$ & $\surd$ & $\surd$ & $\surd$ \\
$F_{88}$ &  &  & $\sum_{(\Delta_i,\Delta_j)_o}\theta[\Delta_i]^8\theta[\Delta_j]^8$ & $\surd$ & $\surd$ \\
$F_{8}$ &  &  &  & $(\sum_{\Delta}\theta[\Delta]^8)^2$ & $\surd$ \\
$G_3[0]$ &  &  &  & $G_3[0]$ & $\surd$ \\
$G_4[0]$ &  &  &  &  & $G_4[0]$ \\
\bottomrule
\end{tabular}
\caption{Basis for the $O^+-$invariants}
\label{tab:Obasis}
\end{center}
\end{table}
We will indicate generically with $e_i^{(g)}$ the elements of the genus $g$ basis. Each function in Table \ref{tab:Obasis} is a suitable polynomial
of degree sixteen in the theta constants and the forms
$\Xi_8^{(g)}[0^{(g)}]$ are suitable linear combinations of them. In order to
compare the two expressions of the proposed superstring chiral measure for $g\leq5$ we also need the Fourier coefficients of the basis of the
$O^+$-invariants. In general, given two series $\sum_n a_nq^n$ and
$\sum_m b_mq^m$, their product is $\sum_n a_nq^n\sum_m
b_mq^m=\sum_kc_kq^k$, with $c_k=\sum_{m+n=k}a_nb_m$. In this way one computes the Fourier
coefficients of the eight functions starting from the ones of the theta constants. However, for increasing $g$
the computation becomes extremely lengthy, due to the huge number of monomials appearing in the definition of the
$e_i^{(g)}$. Thus, although in principle possible by hand, we perform the computation using the \texttt{C++} programming language, see
appendix \ref{program}.

\section{CDG ans\"atze and OPSY ans\"atze for $\Xi_8^{(g)}$} \label{ansatze} 
Before starting the computation of the Fourier coefficients we review the expressions of the forms $\Xi_8^{(g)}[0^{(g)}]$ for $g\leq 5$ in
both formalisms. In what follows we will call $\Xi_8^{(g)}[0]_{CDG}$ the forms defined in \cite{CDG,CDG2,D} (even though
we will use the particular basis of \cite{D}) and $\Xi_8^{(g)}[0]_{OPSY}$ the forms of \cite{OPSY}.
The expressions of the forms $\Xi_8^{(g)}[0^{(g)}]$ found in \cite{D} using the classical theta constants,
see also \cite{CDG, CDG2} for the case $g\leq 4$, are:
\begin{align*}
\Xi_8^{(1)}[0]_{CDG}&=\frac{2}{3}F_1^{(1)}-\frac{1}{3}F_2^{(1)},\\
\Xi_8^{(2)}[0]_{CDG}&=\frac{2}{3}F_1^{(2)}+\frac{1}{3}F_2^{(2)}-\frac{1}{2}F_3^{(2)},\\
\Xi_8^{(3)}[0]_{CDG}&=\frac{1}{3}F_1^{(3)}+\frac{1}{3}F_2^{(3)}-\frac{1}{4}F_3^{(3)}-\frac{1}{64}F_8^{(3)}+\frac{1}{16}F_{88}^{(3)},\\
\Xi_8^{(4)}[0]_{CDG}&=\frac{1}{6}F_1^{(4)}+\frac{1}{3}F_2^{(4)}-\frac{1}{8}F_3^{(4)}+\frac{1}{64}F_8^{(4)}-\frac{1}{16}F_{88}^{(4)}-\frac
12G_3^{(4)}[0^{(4)}]-c_4J^{(4)},\\
\Xi_8^{(5)}[0]_{CDG}&=
\frac{1}{12}F_1^{(5)}+\frac{1}{3}F_2^{(5)}-\frac{1}{16}F_3^{(5)}-\frac{1}{32}F_8^{(5)}+\frac{1}{8}F_{88}^{(5)}-\frac14G_3^{(5)}[0^{(5)}]
+ 2G_4^{(5)}[0^{(5)}]-c_5J^{(5)}.
\end{align*}
Here we have included the terms $-c_4J^{(4)}$ and $-c_5J^{(5)}$ to have vanishing
cosmological constant on the whole $\hh_4$ and $\hh_5$ and to compare
these functions to the ones of \cite{OPSY}. In particular, $c_4=\frac{3^2\cdot
  5}{2^6\cdot 7\cdot 17}$ and $c_5=\frac{17}{2^5\cdot 7 \cdot 11}$ (see \cite{D}).
The forms $\Xi_8^{(g)}[0^{(g)}]$ defined in \cite{OPSY} by means of the
lattice theta series are:
\begin{align*}
\Xi_8^{(1)}[0]_{OPSY}&=-\frac{31}{32}\Theta_{(D_8\oplus D_8)^+}+\frac{512}{315}\Theta_{\zz\oplus A_{15}^+}-\frac{16}{21}\Theta_{\zz^2\oplus (E_7 \oplus E_7)^+}+\frac{1}{9}\Theta_{\zz^4\oplus D_{12}^+}-\frac{1}{168}\Theta_{\zz^8\oplus E_8}\\
&+\frac{1}{10080}\Theta_{\zz^{16}},\\
\Xi_8^{(2)}[0]_{OPSY}&=\frac{155}{512}\Theta_{(D_8\oplus D_8)^+}-\frac{16}{21}\Theta_{\zz\oplus A_{15}^+}+\frac{23}{42}\Theta_{\zz^2\oplus (E_7 \oplus E_7)^+}-\frac{3}{32}\Theta_{\zz^4\oplus D_{12}^+}+\frac{29}{5376}\Theta_{\zz^8\oplus E_8}\\
&-\frac{1}{10752}\Theta_{\zz^{16}},\\
\Xi_8^{(3)}[0]_{OPSY}&=-\frac{155}{4096}\Theta_{(D_8\oplus D_8)^+}+\frac{1}{9}\Theta_{\zz\oplus A_{15}^+}-\frac{3}{32}\Theta_{\zz^2\oplus (E_7 \oplus E_7)^+}+\frac{101}{4608}\Theta_{\zz^4\oplus D_{12}^+}-\frac{3}{2048}\Theta_{\zz^8\oplus E_8}\\
&+\frac{1}{36864}\Theta_{\zz^{16}},\\
\Xi_8^{(4)}[0]_{OPSY}&=\frac{31}{16384}\Theta_{(D_8\oplus
  D_8)^+}-\frac{1}{168}\Theta_{\zz\oplus
  A_{15}^+}+\frac{29}{5376}\Theta_{\zz^2\oplus (E_7 \oplus
  E_7)^+}-\frac{3}{2048}\Theta_{\zz^4\oplus
  D_{12}^+}\\
  &+\frac{23}{172032}\Theta_{\zz^8\oplus E_8}-\frac{1}{344064}\Theta_{\zz^{16}}-b_4 \left(\Theta_{E_8\oplus E_8}-\Theta_{D_{16}^+}\right),\\
\Xi_8^{(5)}[0]_{OPSY}&=-\frac{1}{32768}\Theta_{(D_8\oplus D_8)^+}+\frac{1}{10080}\Theta_{\zz\oplus A_{15}^+}-\frac{1}{10752}\Theta_{\zz^2\oplus (E_7 \oplus E_7)^+}+\frac{1}{36864}\Theta_{\zz^4\oplus D_{12}^+}\\
&-\frac{1}{344064}\Theta_{\zz^8\oplus E_8}
+\frac{1}{10321920}\Theta_{\zz^{16}}-b_5 \left(\Theta_{E_8\oplus E_8}-\Theta_{D_{16}^+}\right).
\end{align*}
Here $b_4=\frac{2^2\cdot 3^3\cdot 5\cdot 11}{7\cdot 17}$ and
$b_5=-\frac{2^5\cdot 17}{7\cdot 11}$ (see \cite{MV}) make the
cosmological constant vanishing on the whole $\hh_4$ and $\hh_5$ respectively.
One of the goals of this paper is to show that up to genus five the two
expressions for the superstring chiral measure coincide. For $g\leq 4$ this was expected
from the uniqueness theorems proved in \cite{CDG,DG,OPSY}. Instead,
for $g=5$ the formalism of the classical theta constants and the one
of the lattice theta series lead to distinct functions both
satisfying the three constraints of Section \ref{ssmeas}. Actually, this indetermination  could appear for each choice for the basis of the spaces $M_8^\theta(\Gamma_5(2))^{O^+}$ or $M_8^{\Theta_S}(\Gamma_5(2))$. Moreover,
their difference is proportional to the Schottky form $J^{(5)}$ and the two forms become equivalent if one requires also the vanishing of the cosmological constants, i.e. the vanishing of their sum over all the even characteristics, $\sum_\Delta \Xi_8^{(g)}[\Delta^{(g)}]=0$. The forms $\Xi_8^{(g)}[\Delta^{(g)}]$ are obtained from the $\Xi_8^{(g)}[0^{(g)}]$ by the action of the symplectic group, see \cite{CDG}.

\section{Change of basis}
In this section we search the relations between the functions defined in Section \ref{rtc} and the lattice theta series. For $g\leq 3$ one can
proceed in several way, but for $g\geq 4$ the knowledge of the Fourier coefficients becomes necessary.

\subsection{The case g=1}\label{g1}
In genus one we can expand the eight lattice theta series on the basis of $O^+$-invariants $F_1^{(1)}$, $F_2^{(1)}$, $F_{16}^{(1)}$ using the
Table 2 in \cite{OPSY}, page 491, that we reproduce in Table \ref{tab:opsy}.
\begin{table}[!h]
\begin{center}
\begin{tabular}{llccc}
\toprule
$i$ & $\Lambda_i$ & $\tau_i$ & $b_i$ & $c_i$ \\
\midrule
0 & $(D_8\oplus D_8)^+$                      & 0 & 1  & 0\\
1 & $\zz\oplus A_{15}^+$                      & 2 & 1  & 0 \\
2 & $\zz^2\oplus (E_7 \oplus E_7)^+$  & 4 & 1  & 0 \\
3 & $\zz^4\oplus D_{12}^+$                  & 8 & 1  & 0 \\
4 & $\zz^8\oplus E_8$                          & 16 & 1  & 0\\
5 & $\zz^{16}$                                       & 32 & 1  & 0\\
6 & $E_8\oplus E_8$                              & 0 & 0  & 1 \\
7 & $D_{16}^+$                                       & 0 & 0  & 1 \\
\bottomrule
\end{tabular}
\caption{Linear relation between lattice theta series.}
\label{tab:opsy}
\end{center}
\end{table}
There, $\Lambda_i$, $i=0,\cdots,7$ label the eight lattices and $\tau_i$, $b_i$ and $c_i$ are the coefficients of the linear expansions of
the series $\Theta_{\Lambda_i}$ on the basis $\Xi_8^{(1)}[0^{(1)}]_{OPSY}$, $\Theta_{\Lambda_0}^{(1)}$, $\Theta_{\Lambda_6}^{(1)}$ for the
space $[\Gamma_1(1,2),8]$. Thus, $\Theta_{\Lambda_i}^{(1)}=\tau_i\Xi_8^{(1)}[0^{(1)}]+b_i \Theta_{\Lambda_0}^{(1)}+c_i \Theta_{\Lambda_6}^{(1)}$.
It is easy to show that the relations
$\Xi_8^{(1)}[0^{(1)}]_{OPSY}=\frac{1}{16}\theta[{}_0^0]^4\eta^{12}=\frac{1}{24}F_1^{(1)}-\frac{1}{48}F_2^{(1)}$
(cf. \cite{D}, section 4.1), $\Theta_{\zz^{16}}=\theta[{}_0^0]^{16}\equiv F_1^{(1)}$
(cf. \cite{CS}, first formula, page 46), $\Theta_{(D_8\oplus D_8)^+}=-\frac 13
F_1^{(1)}+\frac 23 F_2^{(1)}$ (by the fifth line of
Table \ref{tab:opsy}) and $\Theta_{E_8\oplus E_8}=\frac 12
F_{16}^{(1)}$ (cf. \cite{CS}, last formula, page 47) hold. Thus,
the linear
relations of Table \ref{tab:expg1} follow immediately.
\begin{table}[!h]
\begin{center}
\begin{tabular}{lccc}
\toprule
Theta series/Basis &
$F_1$ & $F_2$ & $F_{16}$ \\
\midrule
$\Theta_{(D_8\oplus D_8)^+}$ & -1/3 & 2/3  & 0\\
$\Theta_{\zz\oplus A_{15}^+}$ & -1/4 & 15/24  & 0 \\
$\Theta_{\zz^2\oplus (E_7 \oplus E_7)^+}$ & -1/6 & 7/12  & 0 \\
$\Theta_{\zz^4\oplus D_{12}^+}$ & 0 & 1/2  & 0 \\
$\Theta_{\zz^8\oplus E_8}$ & 1/3 & 1/3  & 0\\
$\Theta_{\zz^{16}}$ & 1 & 0  & 0\\
$\Theta_{E_8\oplus E_8}$ & 0 & 0  & 1/2 \\
$\Theta_{D_{16}^+}$ & 0 & 0  & 1/2 \\
\bottomrule
\end{tabular}
\caption{Theta series on the basis $F_1$, $F_2$ and $F_{16}$.}
\label{tab:expg1}
\end{center}
\end{table}

Moreover, the lattice theta series in genus one are not all linear independent, but they generate a three dimensional vector
space. Therefore, they must satisfy some linear relations, which can
be obtained studying the five dimensional kernel of the first three bold columns of Table \ref{tab:fouriercoef} computed with Magma.
This give the following relations among theta series:
\bes
\begin{pmatrix}
0 & 0 & 0 & 0 & 0 & 0 & -1 & 1 \\
15 & -16 & 0 & 0 & 0 & 1 & 0 & 0 \\
7 & -8 & 0 & 0 & 1 & 0 & 0 & 0 \\
3 & -4 & 0 & 1 & 0 & 0 & 0 & 0 \\
1 & -2 & 1 & 0 & 0 & 0 & 0 & 0 \\
\end{pmatrix}
\begin{pmatrix}
\Theta_{(D_8\oplus D_8)^+} \\
\Theta_{\zz\oplus A_{15}^+} \\
\Theta_{\zz^2\oplus (E_7 \oplus E_7)^+} \\
\Theta_{\zz^4\oplus D_{12}^+} \\
\Theta_{\zz^8\oplus E_8} \\
\Theta_{\zz^{16}} \\
\Theta_{E_8\oplus E_8} \\
\Theta_{D_{16}^+}
\end{pmatrix}
=
\underline{0}.
\ees
As a check, one can show that these relations are in complete agreement with those that can be computed using Table \ref{tab:expg1}.
For example, from the second line one reads $15 \Theta_{(D_8\oplus D_8)^+} -16 \Theta_{\zz\oplus A_{15}^+} + \Theta_{\zz^8\oplus E_8}=0$.
From the Fourier coefficients of the eight theta series and from their
expansion on the basis of the $O^+-$invariants we can also find the
Fourier coefficients for the three functions $F_1^{(1)}$, $F_2^{(1)}$ and $F_{16}^{(1)}$
expressed as polynomials of degree sixteen in the classical theta
constants as showed in Table \ref{tab:fourierFg1}. As this space is
three dimensional, we just need three coefficients and we
choose the ones corresponding to the matrices (that in $g=1$ are just numbers)
$1$, $2$ and $3$. Using the \texttt{C++} program
(cf. Appendix \ref{program}) we also checked the correctness of the
coefficients and further we computed the coefficient corresponding to
the matrix $0$. Actually for lower genus this computation can be performed easily by hand.

\begin{table}[hbtp]
\begin{center}
\begin{tabular}{ccccc}
\toprule
Functions/$m$&$ 0 $ & $ {\bf 1} $&$ {\bf 2} $&$ {\bf 3}$ \\
\midrule
$ F_1 $  & 1 & 32    &  480  & 4480   \\
$ F_2$  & 2 & 16 & 576 & 8384 \\
$ F_{16}$ & 2 & 0 & 960 & 0 \\
\bottomrule
\end{tabular}
\caption{Fourier coefficients for the $F_1$, $F_2$ and $F_{16}$ in genus one.}
\label{tab:fourierFg1}
\end{center}
\end{table}

\subsection{The case g=2}
Using the factorization properties of the classical theta constants one obtains the factorization of the basis of the space of $O^{(+)}$
invariants (cf. \cite{D}, section 4.2), whereas for the theta series one can apply property \eqref{factts}. Thus, we can find the
expansions of the $g=2$ theta series on the basis of the four $O^+-$invariants as follows (sometimes for
brevity we will indicate this space as $O_g$).
In general we have
\be
\Theta^{(g)}_{\Lambda_i}(\tau)=\sum_{j=1}^{\dim O_g}k_{i}^{(g)\,j}e_j^{(g)},
\ee
where $e_j^{(g)}$ are the basis for the genus $g$ $O^+$-invariants, written as polynomials in the classical theta constants, and $k_{i}^{(g)j}$ are
the constants we want to determine.
The restriction on $\hh_1\times \hh_{g-1}$ of the theta series is
\begin{align}\label{spac1}
\Theta^{(g)}_{\Lambda_i}(\tau_{1,g-1})&=\Theta^{(g)}_{\Lambda_i}(\tau_1)\Theta^{(g-1)}_{\Lambda_i}(\tau_{g-1})=\sum_{j=1}^{\dim
O_1}k_{i}^{(1)\,j}e_j^{(1)}\sum_{m=1}^{\dim O_{g-1}}k_{i}^{(g-1)\,m}e_m^{(g-1)}\cr
&=\sum_{j=1}^{\dim
O_1}\sum_{m=1}^{\dim O_{g-1}}k_{i}^{(1)\,j}k_{i}^{(g-1)\,m}e_j^{(1)}e_m^{(g-1)},
\end{align}
but also
\begin{align}\label{spac2}
\Theta^{(g)}_{\Lambda_i}(\tau_{1,g-1})&=\sum_{j=1}^{\dim O_g}k_{i}^{(g)\,j}e_j^{(g)}(\tau_{1,g-1})=
\sum_{j=1}^{\dim O_g}k_{i}^{(g)\,j}(\sum_{l=1}^{\dim
  O_1}a_j^{(1)l}e_l^{(1)}) (\sum_{m=1}^{\dim
  O_{g-1}}a_j^{(g-1)m}e_m^{(g-1)})\cr
&=\sum_{j=1}^{\dim O_g}\sum_{l=1}^{\dim O_1}\sum_{m=1}^{\dim O_{g-1}}k_{i}^{(g)\,j}a_j^{(1)l}a_j^{(g-1)m}e_l^{(1)}e_m^{(g-1)}.
\end{align}
The expressions \eqref{spac1} and \eqref{spac2} must be equal. Thus, for every fixed choice of $l$ and $m$ we obtain a linear
equation in $k_{i}^{(g)\,j}$. The solution of this linear system gives the coefficients in the change of basis.
We give the result for the case $g=2$ in Table \ref{tab:expansiong2}.

\begin{table}[!h]
\begin{center}
\begin{tabular}{lcccc}
\toprule
Theta series/Basis &
$F_1$ & $F_2$ & $F_3$ & $F_{16}$ \\
\midrule
$\Theta_{(D_8\oplus D_8)^+}$ & 1/3 & 2/3  & -1/2 & 0\\
$\Theta_{\zz\oplus A_{15}^+}$ & 7/32 & 35/64  & -45/128 & 0 \\
$\Theta_{\zz^2\oplus (E_7 \oplus E_7)^+}$ & 1/8 & 7/16  & -7/32 & 0 \\
$\Theta_{\zz^4\oplus D_{12}^+}$ & 0 & 1/4  & 0 & 0 \\
$\Theta_{\zz^8\oplus E_8}$ & 0 & 0  & 1/4 & 0 \\
$\Theta_{\zz^{16}}$ & 1 & 0  & 0 & 0\\
$\Theta_{E_8\oplus E_8}$ & 0 & 0 & 0 & 1/4 \\
$\Theta_{D_{16}^+}$ & 0 & 0  & 0 & 1/4 \\
\bottomrule
\end{tabular}
\caption{Theta series on the basis $F_1$, $F_2$, $F_3$ and $F_{16}$.}
\label{tab:expansiong2}
\end{center}
\end{table}
As expected (cf. \cite{DG,D,OPSY}), the matrix of the coefficients has rank four, which is then also the dimension of the kernel
and we can determine the linear relations among the theta series
\bes
\begin{pmatrix}
0 & 0 & 0 & 0 & 0 & 0 & -1 & 1 \\
-105 & 224 & -120 & 0 & 0 & 1 & 0 & 0 \\
-21 & 48 & -28 & 0 & 1 & 0 & 0 & 0 \\
-3 & 8 & -6 & 1 & 0 & 0 & 0 & 0
\end{pmatrix}
\begin{pmatrix}
\Theta_{(D_8\oplus D_8)^+} \\
\Theta_{\zz\oplus A_{15}^+} \\
\Theta_{\zz^2\oplus (E_7 \oplus E_7)^+} \\
\Theta_{\zz^4\oplus D_{12}^+} \\
\Theta_{\zz^8\oplus E_8} \\
\Theta_{\zz^{16}} \\
\Theta_{E_8\oplus E_8} \\
\Theta_{D_{16}^+}
\end{pmatrix}
=
\underline{0}.
\ees
For example, from the third line, we have $-21\ \Theta_{(D_8\oplus D_8)^+}+48\ \Theta_{\zz\oplus A_{15}^+}-28\ \Theta_{\zz^2\oplus
(E_7 \oplus E_7)^+} + \Theta_{\zz^8\oplus E_8}=0$.
One can verify that the same relations result by the study of the kernel of the first four bold columns of the Table \ref{tab:fouriercoef} of
the Fourier coefficients for the lattice theta series.

As for the genus one case, we compute the Fourier coefficients for the
four functions $F_1^{(2)}$, $F_2^{(2)}$, $F_3^{(2)}$ and $F_{16}^{(2)}$ both using the
previous results and the \texttt{C++} program. The Table \ref{tab:fourierFg2} shows the result.
\begin{table}[hbtp]
\begin{center}
\begin{tabular}{ccccccc}
\toprule
Functions/m & $(0,0)$ & ${\bf (1,0)} $&${\bf (2,0)} $&${\bf (3,0)}$ &$
{\bf (1,1)}$ &$ (2,2)$ \\
\midrule
$ {\bf F_1} $ & 1  & 32    &  480  & 4480 & 960 & 175680  \\
$ {\bf F_2} $  & 4 & 32 & 1152 & 16768 & 192 & 243456\\
$ {\bf F_3} $ & 4 & 64 & 1408 & 17152 & 896 & 363776 \\
$ {\bf F_{16}} $ & 4 & 0 & 1920 & 0 & 0 & 702720 \\
$ F_8$ & 16 & 0 & 7680 & 0 & 0 & 2810880 \\
$F_{88}$ & 0 & 0 & 1024 & -16384 & 0 & 546816 \\
\bottomrule
\end{tabular}
\caption{Fourier coefficients for the $F_1$, $F_2$, $F_3$ and $F_{16}$ in genus two.}
\label{tab:fourierFg2}
\end{center}
\end{table}

\subsection{The case $g=3$}
In genus three we can obtain the expansion of the theta series on the
basis $e_i^{(3)}$ with the method of factorization explained in the
previous section. We report the result in Table \ref{tab:expansiong3}.
\begin{table}[!h]
\begin{center}
\begin{tabular}{lccccc}
\toprule
Theta series/Basis &
$F_1$ & $F_2$ & $F_3$ & $F_{16}$ & $F_{88}$ \\
\midrule
$\Theta_{(D_8\oplus D_8)^+}$ & 0 & 0  & 0 & 1/8 & -1/16\\
$\Theta_{\zz\oplus A_{15}^+}$ & 7/512 & 35/512  & -45/2048 & 315/4096 & -315/8192 \\
$\Theta_{\zz^2\oplus (E_7 \oplus E_7)^+}$ & 1/64 & 7/64  & -7/256 & 21/512 & -21/1024 \\
$\Theta_{\zz^4\oplus D_{12}^+}$ & 0 & 1/8  & 0 & 0 & 0\\
$\Theta_{\zz^8\oplus E_8}$ & 0 & 0  & 1/8 & 0 & 0 \\
$\Theta_{\zz^{16}}$ & 1 & 0  & 0 & 0 & 0\\
$\Theta_{E_8\oplus E_8}$ & 0 & 0  & 0 & 1/8 & 0 \\
$\Theta_{D_{16}^+}$ & 0 & 0  & 0 & 1/8 & 0 \\
\bottomrule
\end{tabular}
\caption{Theta series on the basis $F_1$, $F_2$, $F_3$, $F_{16}$ and
  $F_{88}$ in genus three.}
\label{tab:expansiong3}
\end{center}
\end{table}
As expected, the matrix of the coefficients has rank five, thus its kernel has dimension three.
Again we find the linear relations studying the kernel of the matrix:
\bes
\begin{pmatrix}
0 & 0 & 0 & 0 & 0 & 0 & -1 & 1 \\
315 & -896 & 720 & -140 & 0 & 1 & 0 & 0 \\
21 & -64 & 56 & -14 & 1 & 0 & 0 & 0 \\
\end{pmatrix}
\begin{pmatrix}
\Theta_{(D_8\oplus D_8)^+} \\
\Theta_{\zz\oplus A_{15}^+} \\
\Theta_{\zz^2\oplus (E_7 \oplus E_7)^+} \\
\Theta_{\zz^4\oplus D_{12}^+} \\
\Theta_{\zz^8\oplus E_8} \\
\Theta_{\zz^{16}} \\
\Theta_{E_8\oplus E_8} \\
\Theta_{D_{16}^+}
\end{pmatrix}
=
\underline{0}.
\ees
As in the two previous cases, the same linear relations follow from the
Table \ref{tab:fouriercoef} of the Fourier coefficients of the lattice
theta series considering the first five bold columns.

As for genus one and two we compute the Fourier coefficients for the
functions $F_1^{(3)}$, $F_2^{(3)}$, $F_3^{(3)}$, $F_{16}^{(3)}$ and $F_{88}^{(3)}$ and we control
the result using the computer. In Table \ref{tab:fourierFg3} we show the result.
\begin{table}[hbtp]
\begin{center}
\resizebox*{1\textwidth}{!}{
\begin{tabular}{ccccccccc}
\toprule
Functions/m& $(0,0,0)$ & ${\bf (1,0,0)} $&${\bf (2,0,0)} $&${\bf (3,0,0)}$ &$
{\bf (1,1,0)}$ &$ (2,2,0)$ & ${\bf (1,1,1)}$ & $(2,2,2)$ \\
\midrule
$ {\bf F_1} $ &1  & 32    &  480  & 4480 & 960 & 175680 & 26880 &  47174400 \\
$ {\bf F_2}$   &8 & 64 & 2304 & 33536 & 384 & 486912 & 1536  & 73451520\\
$ {\bf F_3}$   &8 & 128 & 2816 & 34304 & 1792 & 727552 & 21504 & 137410560\\
$ {\bf F_{16}}$ &8 & 0 & 3840 & 0 & 0 & 1405440 & 0 & 377395200\\
$ F_{8}$  &64 & 0 & 30720 & 0 & 0 & 11243520 & 0 & 3019161600\\
$ {\bf F_{88}}$ &0 & 0 & 4096 & -65536 & 0 & 2187264 & 0 & 673677312\\
$G_3[0]$ & 1 & 0 & 224 & 4096 & 0 & 38976 & 0 & 5069568  \\
\bottomrule
\end{tabular}
}
\caption{Fourier coefficients for the $F_1$, $F_2$, $F_3$, $F_{16}$
  and $F_{88}$ in genus three.}
\label{tab:fourierFg3}
\end{center}
\end{table}
We also compute the Fourier coefficients of the functions $F_8^{(3)}$
and $G_3^{(3)}[0^{(3)}]$. Thus, we get another proof of the relation (cf. \cite{D}, page 20):
\be
G_3^{(3)}[0^{(3)}]=\frac{1}{64}F_8^{(3)}-\frac{1}{16}F_{88}^{(3)}-\frac{5}{448}(8F_{16}^{(3)}-F_8^{(3)}),
\ee
as can be check inserting in the previous equation the Fourier coefficients.

\subsection{The case $g=4$} \label{g4}
The genus four case is the first interesting case because the factorization approach does no more work. The failure of this method is
due to the fact that the space of moduli of curves is not the whole Siegel upper half plane. Indeed, the two theta series defined by the
lattice $D_{16}^+$ and $E_8\oplus E_8$ are no longer the same function and the differences among this two functions are lost by restricting
on the boundary of $\hh_4$.

Thus, in order to find the relations between the lattice theta series and the functions $e_i^{(4)}$ we need the Fourier coefficients of
the functions $e_i^{(4)}$. We have computed them with the \texttt{C++} program. The result are reported in Table \ref{tab:fourierFg4}.
\begin{table}[hbtp]
\begin{center}
\resizebox*{1\textwidth}{!}{
\begin{tabular}{ccccccccccc}
\toprule
&${\bf (0,0,0,0,0)}$ & $ {\bf  (1,0,0,0,0)} $&$ {\bf  (2,0,0,0,0)} $&$ {\bf  (3,0,0,0,0)}$ &$
{\bf  (1,1,0,0,0)}$ &$ (2,2,0,0,0)$ & ${\bf  (1,1,1,0,0)}$ & $(2,2,2,0,0)$ & ${\bf  (1,1,1,1,0)}$ & ${\bf  (2,2,2,2,0)}$\\
\midrule
$ {\bf F_1} $   &1& 32    &  480  & 4480 & 960 & 175680 & 26880 & 47174400 & 698880 & 8858304000\\
${\bf F_2}$    &16 &128 & 4608 & 67072 & 768 & 973824 & 3072 & 146903040 & 6144 & 15427215360 \\
${\bf F_3}$    &16 & 256 & 5632 & 68608 & 3584 & 1455104 & 43008 &
274821120 & 430080 & 37058273280 \\ 
${\bf F_{16}}$ &16& 0 & 7680 & 0 & 0 & 2810880 & 0 & 754790400 & 0 &  141732864000 \\
${\bf F_{8}}$  & 256 & 0 & 122880 & 0 & 0 & 44974080 & 0 & 12076646400 & 0 & 2320574054400 \\
${\bf F_{88}}$ & 0 & 0 & 16384 & -262144 & 0 & 8749056 & 0 & 2694709248 & 0
& 549726191616 \\
${\bf G_3[0]} $  & 15  & 32 & 3616 & 61824 & -64 &
655808 & 256 & 85511424 & -1536 & 8099185152\\
$G_4[0] $ & 1 & 0 & 224 & 4096 & 0 & 38976 & 0 & 5069568 & 0 &
386797056 \\
$J^{(4)}$ & 0 & 0 & 0 & 0 & 0 & 0 & 0 & 0 & 0 & -52848230400 \\
\bottomrule
\end{tabular}
}
\caption{Fourier coefficients for the basis $F_1$, $F_2$, $F_3$, $F_{16}$, $F_{88}$, $F_8$ and $G_3[0]$ in genus four. In addition we
compute the coefficients of $G_4[0]$ and of $J^{(4)}$.}
\label{tab:fourierFg4}
\end{center}
\end{table}
Adding the rows of this table to the ones of Table \ref{tab:fouriercoef} and considering the first seven bold columns, one finds, as
expected, that the complete matrix has rank seven. Again, we get the
expansions of the lattice theta series on the basis $e_i^{(4)}$. The result is shown in Table \ref{tab:expansiong4}.
\begin{table}[!h]
\begin{center}
\resizebox*{1\textwidth}{!}{
\begin{tabular}{lccccccc}
\toprule
Theta series/Basis &
$F_1$ & $F_2$ & $F_3$ & $F_{16}$ & $F_{88}$ & $F_8$ & $G_3[0]$\\
\midrule
$\Theta_{(D_8\oplus D_8)^+}$ & 0 & 0  & 0 & 0 & -1/64 &
1/256 & 0 \\
$\Theta_{\zz\oplus A_{15}^+}$ & 7/8192 & 35/4096  & -45/32768 & 135/16384 & -315/65536 &
45/65536 & 315/8192 \\
$\Theta_{\zz^2\oplus (E_7 \oplus E_7)^+}$ & 1/512 & 7/256  & -7/2048
& 0 & 0& 0& 21/512 \\
$\Theta_{\zz^4\oplus D_{12}^+}$ & 0 & 1/16  & 0 & 0 & 0 & 0 & 0\\
$\Theta_{\zz^8\oplus E_8}$ & 0 & 0  & 1/16 & 0 & 0 & 0 & 0\\
$\Theta_{\zz^{16}}$ & 1 & 0  & 0 & 0 & 0 & 0 & 0\\
$\Theta_{E_8\oplus E_8}$ & 0 & 0  & 0 & 0 & 0 & 1/256 & 0\\
$\Theta_{D_{16}^+}$ & 0 & 0  & 0 & 1/16 & 0 & 0 & 0\\
\bottomrule
\end{tabular}
}
\caption{Theta series on the basis $F_1$, $F_2$, $F_3$, $F_{16}$, $F_{88}$, $F_8$ and $G_3[0]$ in genus four.}
\label{tab:expansiong4}
\end{center}
\end{table}
These Fourier coefficients also provide a proof of the relation (cf. \cite{D}, page 23):
\begin{align*}
G_4^{(4)}[0^{(4)}]&=\frac{1}{256}F_8^{(4)}-\frac{1}{64}F_{88}^{(4)}+\frac{3}{1792}J^{(4)}\\
&=\frac{1}{448}F_8^{(4)}-\frac{1}{64}F_{88}^{(4)}+\frac{3}{112}F_{16}^{(4)}.
\end{align*}
Moreover, we obtain a linear relation between the lattice theta series
\bes
\begin{pmatrix}
1 & -1024/315 & 64/21 & -8/9 & 2/21 & -1/315 & -3/7 & 3/7 \\
\end{pmatrix}
\begin{pmatrix}
\Theta_{(D_8\oplus D_8)^+} \\
\Theta_{\zz\oplus A_{15}^+} \\
\Theta_{\zz^2\oplus (E_7 \oplus E_7)^+} \\
\Theta_{\zz^4\oplus D_{12}^+} \\
\Theta_{\zz^8\oplus E_8} \\
\Theta_{\zz^{16}} \\
\Theta_{E_8\oplus E_8} \\
\Theta_{D_{16}^+}
\end{pmatrix}
=
\underline{0}.
\ees

\subsection{The case $g=5$} \label{caseg5}
In genus five, we consider the eight columns of Table \ref{tab:fouriercoef}. This matrix has rank eight, so all the theta
series are linearly independent. As in genus four, to study the relations between the Riemann theta constants and the lattice theta series we
need the Fourier coefficients of the functions $e_i^{(5)}$. We have computed them by the computer and we report the result in Table \ref{tab:fourierFg5}, that also has rank eight.
\begin{table}[hbtp]
\begin{center}
\resizebox*{1\textwidth}{!}{
\begin{tabular}{cccccccccccc}
\toprule
&${\bf (0,0,0,0,0)}$ & $ {\bf  (1,0,0,0,0)} $&$ {\bf  (2,0,0,0,0)} $&$ {\bf  (3,0,0,0,0)}$ &$
{\bf  (1,1,0,0,0)}$ &$ (2,2,0,0,0)$ & ${\bf  (1,1,1,0,0)}$ &
$(2,2,2,0,0)$ & ${\bf  (1,1,1,1,0)}$ & ${\bf  (2,2,2,2,0)}$  & ${\bf
  (1,1,1,1,1)}$ \\
\midrule
${\bf F_1} $   &1& 32 & 480  & 4480 & 960 & 175680 & 26880 & 47174400 &698880 & 8858304000 & 16773120  \\
${\bf  F_2}$    &32 &256 & 9216 & 134144 & 1536 & 1947648 & 6144 & 293806080
& 12288 &30854430720 & 0 \\
$ {\bf F_3}$    &32 & 512 & 11264 & 137216 & 7168 & 2910208 & 86016 &
549642240 & 860160 &
74116546560 & 6881280\\
$ {\bf F_{16}}$ &32& 0 & 15360 & 0 & 0 & 5621760 & 0 & 1509580800 & 0 & 283465728000 & 0\\
$ {\bf F_{8}}$  & 1024 & 0 & 491520 & 0 & 0 & 179896320 & 0 & 48306585600 & 0 & 9282296217600 & 0 \\
$ {\bf F_{88}}$ & 0 & 0 & 65536 & -1048576 & 0 & 34996224 & 0 & 10778836992 & 0 & 2198904766464 &0\\
$ {\bf G_3[0]} $  & 155& 480& 38560& 640640& 64& 7174336& -2304& 954147072& 22016 & 90356353536 & -225280\\
$ {\bf G_4[0]} $  & 31 & 32 & 7200 & 127360 & -64 & 1279424 & 256 &
166624512 & -1536 & 14287938048 & 12288\\
$J^{(5)}$ &0&0&0&0&0&0&0&0&0&-211392921600&0\\
\bottomrule
\end{tabular}
}
\caption{Fourier coefficients for the $F_1$, $F_2$, $F_3$, $F_{16}$, $F_{88}$, $F_8$, $G_3[0]$ and $G_4[0]$ in genus five.}
\label{tab:fourierFg5}
\end{center}
\end{table}
Gluing this table to the one of the Fourier coefficients for the lattice theta series we obtain a matrix of rank eight. {\em So, all
the lattice theta series can be expressed as linear combination of $e_i^{(5)}$ and vice versa!}
Indeed we can be more precise. As the rank of the whole set of coefficients is $8$, we get 8 linear relations among the two bases:
\begin{eqnarray}
&& F_{16}=2^5 \Theta_{D_{16}^+}, \qquad\  F_8=2^{10} \Theta_{E_8\oplus E_8}, \qquad\  F_1=\Theta_{\zz^{16}}, \label{vere} \\
&& F_3=2^5 \Theta_{\zz^8\oplus E_8}, \qquad\  F_2=2^5 \Theta_{\zz^4\oplus D_{12}^+}, \qquad\
F_8-4F_{88} =2^{10} \Theta_{(D_8\oplus D_8)^+}, \label{q-certe} \\
&& -4F_1-112F_2+7 F_3 -84 G_3 =-16384 \Theta_{\zz^2\oplus(E_7\oplus E_7)^+}, \label{suspect1} \\
&& -28 F_1 -560 F_2 +45F_3-1260 G_3 -10080G_4=-524288 \Theta_{\zz\oplus A_{15}^+} \label{suspect2}.
\end{eqnarray}
Note that the relations
\eqref{vere} can be directly checked. The relations \eqref{q-certe} also are simply a generalization of the lower genus ones.
However, for all the relations we can also give some consistency cheks. Summing each side of the eight equalities over the 528 characteristics we obtain eight identities. For example for the \eqref{suspect2} we obtain $-524288F_{16}^{(5)}=-524288\cdot 2^5\Theta_{D_{16}^+}$ and $2^5\Theta_{D_{16}^+}$ is exactly the $F_{16}^{(5)}$. These sums can be performed using Table 6 of \cite{D} and Table 1 and Appendix B.2
of \cite{MV}.
Moreover, one verifies that also the restriction to $\hh_1\times \hh_4$ of each equality is an identity.
\section{Equivalence of the CDG and the OPSY construction} \label{equivalence}
In this section we prove the equivalence of the two functions constructed using the classical theta functions and the lattice theta series. They at most differ by a multiple of the Schottky form and become identical if one fixes the value of the cosmological constant to zero. We first study the Fourier coefficients of the two $\Xi_8^{(g)}[0^{(g)}]$, then we give an analytic proof of their equivalence.
\subsection{Fourier coefficients for the partition function}
Inserting  the Fourier coefficients of the basis $e_i^{(g)}$ and of the lattice theta series in the definition of the functions
$\Xi_8^{(g)}[0^{(g)}]$ of Section \ref{ansatze} we can compute, for every genus $g\leq 5$, the Fourier expansions of the $\Xi_8^{(g)}[0^{(g)}]$.
Table \ref{tab:compare} shows these coefficients for the two expressions of the forms $\Xi_8$. We also add $0$ in the first column for the functions $ \Xi_8^{(g)}[0]_{OPSY} $, because, from the geometric discussion of Section \ref{lts}, it is clear that there are no vectors in the lattice of null norm.
\begin{table}[hbtp]
\begin{center}
\resizebox*{1\textwidth}{!}{
\begin{tabular}{lcccccccccccc}
\toprule
& ${\bf (0,0,0,0,0)}$ & ${\bf (1,0,0,0,0)} $&${\bf (2,0,0,0,0)} $&${\bf (3,0,0,0,0)}$ &$
{\bf (1,1,0,0,0)}$ &$ (2,2,0,0,0)$ & ${\bf (1,1,1,0,0)}$ & $(2,2,2,0,0)$ & ${\bf (1,1,1,1,0)}$ & ${\bf (2,2,2,2,0)}$ &${\bf (1,1,1,1,1)}$\\
\midrule
$ \Xi_8^{(1)}[0]_{OPSY} $   & 0 & 1 & 8  & 12 &  &  & &&&&   \\[1em]
$ \Xi_8^{(1)}[0]_{CDG} $                    & 0 & 16 & 128 & 192 &&&&&&&\\
\midrule
$ \Xi_8^{(2)}[0]_{OPSY} $    & 0 &0 & 0 & 0 & 1 & 64 &  &&&&  \\[1em]
$ \Xi_8^{(2)}[0]_{CDG} $                   &0&0&0&0&256& 16384 &&&&&\\
\midrule
$ \Xi_8^{(3)}[0]_{OPSY} $    & 0 &0&0&0&0& 0 & 1 & 192 &&&  \\[1em]
$ \Xi_8^{(3)}[0]_{CDG} $                    &0&0&0&0&0&0&4096 &786432&&& \\
\midrule
$ \Xi_8^{(4)}[0]_{OPSY}$    & 0 &0& 0 & 0 & 0 & 0 & 0 & 0 & 1 & $\frac{38976}{17}$ & \\[1em]
$ \Xi_8^{(4)}[0]_{CDG} $                    &0&0&0&0&0&0&0&0&65536&$\frac{2554331136}{17}$&\\
\midrule
$ \Xi_8^{(5)}[0]_{OPSY}$    & 0 & 0 & 0 & 0 & 0 & 0 & 0 & 0 & 0 & $\frac{16043183100}{11}$ &1 \\[1em]
$ \Xi_8^{(5)}[0]_{CDG} $  &0&0&0&0&0&0&0&0&0&$\frac{16822496762265600}{11}$&1048576\\
\bottomrule
\end{tabular}
}
\caption{Fourier coefficients for the two expressions of the form $\Xi_8$. In the first line of each genus are the coefficients of the OPSY
forms and in the second line the ones of the CDG forms.}
\label{tab:compare}
\end{center}
\end{table}
We conclude that the two functions are the same up to genus five, apart for an unessential global factor $2^{4g}$ due to the different
definition of the Dedekind function used in \cite{D} and in \cite{OPSY} (cf. footnote 7 in \cite{D}, page 17).

\subsection{Analytic proof of the equivalence of the CDG and the OPSY construction}\label{analproof}
In this section we give an analytic proof of the equivalence of the two constructions of the forms $\Xi_8[\Delta]$ through the study of their restriction to
$\hh_1\times\hh_4$. We will show that $(\Xi_8^{(5)}[0^{(5)}]_{CDG} -\Xi_8^{(5)}[0^{(5)}]_{OPSY})(\tau_{1,4})=0$ on the whole $\hh_1\times \hh_4$. To compare the two expressions of the forms $\Xi_8$ one has to get rid of the factor $2^{4g}$. We choose to multiply $\Xi_8^{(g)}[0]_{OPSY}$ by $2^{4g}$ and that implies that the constants $b_4$ and $b_5$ of Section \ref{ansatze} become $b_4=-\frac{2^7\cdot 3}{7\cdot 17}$ and $b_5=-\frac{2^5\cdot 17}{7\cdot 11}$.
Indeed, using the expressions of Section \ref{ansatze}:
\begin{align}\label{ltspac}
\Xi_8^{(5)}[0^{(5)}]_{OPSY}(\tau_{1,4}) &=\Xi_8^{(1)}[0^{(1)}](\tau_1)
\Xi_8^{(4)}[0^{(4)}](\tau_4)\cr
&+\left(\frac{2^5\cdot 3\cdot 13}{7\cdot 17}\Theta_{\zz^8\oplus E_8}^{(1)}-
\frac{2^6\cdot3^2\cdot 5}{7\cdot 17}\Theta_{\zz_{16}}^{(1)} + \frac{2^5\cdot 17}{7\cdot 11}\Theta_{E_8\oplus E_8}^{(1)}\right)\left(\Theta_{E_8\oplus E_8}^{(4)}-\Theta_{D_{16}^+}^{(4)}\right)\cr
&=\Xi_8^{(1)}[0^{(1)}](\tau_1) \Xi_8^{(4)}[0^{(4)}](\tau_4)\cr
&+\left[\frac{3}{2^2\cdot 7\cdot 17}\left(-\frac 23F_1^{(1)}+ 5 F_2^{(1)}\right)-\frac{17}{2^4\cdot 7\cdot 11}F_{16}^{(1)}\right]J^{(4)},
\end{align}
where we have used the linear relation among the genus four lattice theta series found in \ref{g4}, the genus one relations among the lattice theta series and the basis functions $e_i^{(4)}$ of Section \ref{g1}, and the fact that\footnote{In general $J^{(g)}=-2^{2g}(\Theta_{E_8\oplus   E_8}^{(g)}-\Theta_{D_{16}^+}^{(g)})$.}
$J^{(4)}=-2^8(\Theta_{E_8\oplus E_8}^{(4)}-\Theta_{D_{16}^+}^{(4)})$.
With a similar computation we obtain for the form $\Xi_8^{(5)}[0^{(5)}]_{CDG}$:
\begin{align}
\Xi_8^{(5)}[0^{(5)}]_{CDG}(\tau_{1,4})&= \Xi_8^{(1)}[0^{(1)}](\tau_1) \Xi_8^{(4)}[0^{(4)}](\tau_4) \cr
&+\left(-\frac
{3^2}{2^3\cdot 7\cdot 17}F_1^{(1)}+\frac{3\cdot 5}{2^2\cdot 7 \cdot 17}F_2^{(1)}-\frac{17}{2^4\cdot 7\cdot 11}F_{16}^{(1)}\right)J^{(4)}\cr
&=\Xi_8^{(1)}[0^{(1)}](\tau_1) \Xi_8^{(4)}[0^{(4)}](\tau_4)\cr
&+\left[\frac{3}{2^2\cdot 7\cdot 17}\left(-\frac 23F_1^{(1)}+ 5 F_2^{(1)}\right)-\frac{17}{2^4\cdot 7\cdot 11}F_{16}^{(1)}\right]J^{(4)},
\end{align}
that is exactly the same as \eqref{ltspac}. This and the fact that the sum over the 528 genus five even characteristics of both the forms $\Xi_8^{(5)}[0^{(5)}]$ is a multiple of the Schottky form show the equivalence of the two constructions.
Fixing the value of the cosmological constant and getting rid of the factor $2^{4g}$, they do not differ neither for a multiple of $J^{(5)}$ because, if so, a term proportional to $F_{16}^{(1)}J^{(4)}$ should appear in the difference of their restrictions due to the fact that $J^{(5)}(\tau_{1,4})=2F_{16}^{(1)}J^{(4)}$.
The factorizations can be obtained using the properties of the lattice theta series (see Section \ref{lts}) and the restrictions properties of the functions $e_i^{(5)}$ (see \cite{D}, Section 4.2 and 7.1). Alternatively, one can employs the linear relations found in Section \ref{caseg5}. Indeed changing the basis with those relations one obtains $\Xi_8^{(5)}[0^{(5)}]_{CDG}$ from $\Xi_8^{(5)}[0^{(5)}]_{OPSY}$ and vice versa. This is another check for the computation leading to relations \eqref{vere}, \eqref{q-certe}, \eqref{suspect1} and \eqref{suspect2}.

\section{Conclusions and perspectives}
It is a well known fact (see \cite{atmoosen,atrasen,mooremor,morper,verver}) that the path integral formulation of superstring theory at higher genus is affected by
ambiguities, mainly due to the difficulty in finding a supercovariant formulation. Indeed, even though the super moduli space of super Riemann
surfaces can be locally split in even and odd part, this does not work globally and the result comes out to depend on the choice of
a bosonic slice in a non covariant way (see \cite{CD2} for a review). In a series of papers \cite{DP0,two2,two3,DP}, D'Hoker and Phong have been able to
determine by direct calculation the genus two amplitudes. As a byproduct they formulated a set of ans\"atze that should be satisfied
by the amplitudes at all genera. In \cite{CD} it has been shown that these ans\"atze characterize univocally the genus two measure, but they
required a small modification to work at genus three \cite{CDG,DG}. The same ans\"atze have produced solutions for the genus
four and five cases also \cite{CDG2,D}.

The $g=5$ case is particular, as resulted by the fact that fixing the Schottky term $J^{(5)}$ is necessary to get a vanishing
cosmological constant. Note, however, that this should be a prediction of the theory and not an ansatz \cite{GS,MV}. This has been yet
criticized in \cite{DbMS}. Thus, it becomes unclear if and what modular properties are sufficient to characterize the amplitudes. Such stronger constraints should come out from a more basis formulation of higher genus path integral. The ambiguity left open by the ans\"atze at $g\geq 5$ is an indetermination of the Schottky form contribution. This indetermination can be fixed by requiring also the vanishing of the cosmological constant. Note that the Schottky form in $g=5$ does not vanish on the Jacobi locus \cite{GS}, so the ambiguity is significative even on the locus of curves.

In \cite{OPSY} and \cite{D} the solution of the constraints for $g=5$ has been determined by means of different methods. In the
first paper the authors started from a basis of the lattice theta series of weight eight $M_8^{\Theta_S}(\Gamma_5(2))$,
whereas in the second paper the author starts from a basis of the genus five modular forms of weight eight $M_8^\theta(\Gamma_5(2))^{O^+}$.
In each case it has been determined a unique solution modulo $J^{(5)}$. By computing the Fourier coefficients of both basis here we have
been able to show that this two solutions coincide modulo $J^{(5)}$ and they become exactly the same if we impose the vanishing of the cosmological constant. Thus, we could suspect that a further constraint could imply uniqueness. Moreover, we have shown the complete equivalence
for lower genus and we have completely determined the relations between the corresponding selected basis, then determining an explicit
identification of the spaces $M_8^{\Theta_S}(\Gamma_g(2))$ and $M_8^{\theta}(\Gamma_g(2))^{O^+}$ for $g=1,\ldots,5$.

The previous considerations also lead to the question wether for $g>5$ the ambiguity left open by the three constraints is again an indetermination of the Schottky form contribution or of stronger nature. Moreover, the trick of fixing the value of the cosmological constant does not work for $g>5$, as pointed out in \cite{DbMS}. In addition, for $g\geq 5$, due to the non normality of the ring $M_8(\Gamma_g(2))$, might exist modular forms that are not polynomial in the theta constants satisfying the three constraints. The answer to this kind of questions would lead to a generalization of the uniqueness theorems proved up to genus four.
Moreover, a deeper understanding of the path integral formulation of the theory is now essential to
overcome these problems. However, it is worth to note that we are working here on the whole Siegel space, whereas string quantities
require to be defined on the space of curves only (the Schottky locus).
These points are actually under investigation.

\

\

\section*{Acknowledgments}
We acknowledge Bert van Geemen for numerous discussions and suggestions and for his continuous support.
We also acknowledge Riccardo Salvati-Manni for useful correspondence. We would also like to thank Luca Paredi for giving us the web space for the code.

\begin{appendix}
\section{The program} \label{program}
In this section we briefly present the structure of the program we used to compute the Fourier coefficients of the functions $e_i^{(g)}$. The code is available on \url{http://www.dfm.uninsubria.it/thetac/}

An element  of $\hh_g$ has the generic form:
\be
\tau=
\begin{pmatrix}
\tau_1 & \tau_{g+1} & \cdots& \cdots & \tau_{2g-1} \\
\tau_{g+1} & \tau_2 &  \tau_{2g} & \cdots & \tau_{3g-3} \\
\vdots & \vdots &\ddots&\cdots&\vdots\\
\tau_{2g-2}&\tau_{3g-4}&\cdots& \tau_{g-1}& \tau_{g(g+1)/2}\\
\tau_{2g-1}&\tau_{3g-3}&\cdots&\cdots& \tau_{g}
\end{pmatrix}.
\ee
Thus, from the definition of theta constant \eqref{sum1} it is clear that truncating the series we obtain a polynomial in $g(g+1)/2$ variables $q_{ij}=e^{\pi i\tau_{ij}}$, with $1\leq i\leq j\leq g$ and the same holds true for the functions $e_i^{(g)}$.
It will be useful to rewrite the definition \eqref{sum1} as:
\be \label{theta2}
\theta[{}_a^b](\tau)=(-)^{\frac 12\sum_i a_ib_i}\sum_{m\in\zz^g} (-)^{\sum_i m_ib_i}\left(\prod_i p_{ii}^{(2m_i+a_i)^2}\right)\left(\prod_{i<j} p_{ij}^{2(2m_i+a_i)(2m_j+a_j)}\right),
\ee
with $p_{ij}=q_{ij}^{1/4}$, so the exponents are integer numbers. This renders faster the computations with the computer. The previous expansion may be thought as a polynomial in $p_{ii}$ with coefficients that are polynomials in $p_{ij}$, $i<j$ (this observation will be useful later).

To perform the computation we have defined some \texttt{C++} classes. First, we have defined the generic class \texttt{Polynomial}, defined as \texttt{template <typename CffType, typename ExpType> class Polynomial}, which accepts two types as parameters, \texttt{CffType} and \texttt{ExpType}. \texttt{CffType} represents the type of the coefficient of a single monomial in \texttt{Polynomial} and \texttt{ExpType} the type of the exponent. 
In order to perform the elementary operations with polynomials, we have introduced the operators of addition, multiplication and raising to power for the \texttt{Polynomial} class.
Then, we have defined a simple polynomial with integer coefficients: \texttt{typedef Polynomial<cln::cl\_I, short> IntPol}\footnote{To manage long integer coefficients we use the \texttt{cl\_I} class from CLN library, \url{http://www.ginac.de/CLN/}}. This type will be the coefficient for the \texttt{ThetaPol} polynomial, which will be used to represent the series expansion of the theta constants: \texttt{typedef Polynomial<IntPol, unsigned short> ThetaPol}.

In order to compute the Fourier coefficients corresponding to the ten diagonal matrices of Section \ref{flts} we proceed as follows. For each even theta constant\footnote{Recall that the number of even theta constants is $2^{g-1}(2^{g+1})$.} we ``fill up'' the
\texttt{ThetaPol} by computing the (finite) sums \eqref{theta2} in which each component of $m\in\zz^{g}$ is no bigger than three. Using the operations on the polynomials we just defined, the  \texttt{ThetaPol}'s are the bricks to build up the functions $e_i^{(g)}$ from their definition. Therefore, the function $e_i^{(g)}$ has the generic form:
\be\label{expansion}
e_i^{(g)}(\tau)=\sum_{n_1,\cdots,n_g \in \nn_0}(\cdots)p_{11}^{n_1}\cdots p_{gg}^{n_g},
\ee
where in $(\cdots)$ there are the non diagonal or constant terms. Note that the exponents of the diagonal terms $p_{ii}$ are always positive, hence multiplying the polynomials of the theta constants the exponents cannot decrease. Due to our choice for the ten matrices, we can introduce a sort of ``filter" for the value of the exponents. Roughly speaking, in the expansion \eqref{expansion} we neglect the terms with exponent of $p_{ii}$ ``bigger than the ones appearing in the diagonal of the ten matrices". This allows us to make the computations very fast. Thus, the Fourier coefficients of the matrix $m=\diag (m_1,\cdots,m_g)$ is the constant term in $(\cdots)$ of the monomial with $n_1=4m_1,\cdots,n_g=4m_g$.

\end{appendix}


\end{document}